%% file: MTBLongterm.tex
\documentclass[aps,twocolumn,amsmath,amssymb,a4paper,9pt,showkeys,notitlepage]{revtex4}
\def\pdftitle{Long term}
\def\authorname{Marc Pichel}
\def\pdfsubject{}
\def\pdfkeywords{}
\def\pdfbackref{none}

\include{includes/Header}
\include{includes/Macro}

\begin{document}

\title{Long term observation of \emph{Magnetospirillum
    gryphiswaldense} \\ in a microfluidic channel}
% research question: What does long term observation of
% \emph{Magnetosprilillum Gryphiswaldense} teach us?
\author{T. A. G. Hageman$^{1,2,3*}$}
\author{M. P. Pichel$^{1,2,3}$\footnote{Both authors contributed
    equally to this work}}
\author{P.A. L\"othman$^{1,2,3}$}
\author{J. Cho$^4$}
\author{M. Choi$^4$}
\author{N.Korkmaz$^1$}
\author{Andreas Manz$^{1}$}
\author{L. Abelmann$^{1,2,3}$}
\affiliation{$^{1}$KIST Europe, Saarbr\"ucken, Germany, \\
  $^{2}$University of Twente, Enschede, The Netherlands\\
  $^3$Saarland University, Saarbr\"ucken, Germany\\
  $^4$KBSI, Seoul, Korea\\
\underline{l.abelmann@kist-europe.de}}
%\date{\today}
\vskip10mm
\begin{abstract}
  We controlled and observed individual magnetotactic
  bacteria (\emph{Magnetospirillum gryphiswaldense}) inside a
  \SI{5}{\um} high microfluidic channel for over four
  hours. After a period of constant velocity, the duration of which
  varied between bacteria, all observed bacteria showed a gradual
  decrease in their velocity of about \SI{25}{nm/s^2}. After coming to
  a full stop, different behaviour was observed, ranging from rotation
  around the centre of mass synchronous with the direction of the external magnetic
  field, to being completely immobile. Our results suggest
  that the influence of the high intensity illumination and the
  presence of the channel walls are important parameters to consider
  when performing observations of such long duration.
\end{abstract}

% insert suggested PACS numbers in braces on next line
% insert suggested keywords - APS authors don't need to do this

\maketitle
% \makeatletter 
% \renewcommand{\@tocrmarg}{2.55em plus1fil} 
% \renewcommand{\@pnumwidth}{3em} 
% \renewcommand{\@tocrmarg}{4em} 
% \makeatother

%\tableofcontents

\input{sections/Introduction}
\input{sections/Experimental}
\input{sections/Results}
\input{sections/Discussion}
\input{sections/Conclusion}

\begin{acknowledgments}
  The authors would like to thank Mohammed Elwi Mitwally of the German
  University in Cario for help with the manual control of the MTB.
\end{acknowledgments}

% Make symbolic links to the PaperBase directory. 
% Make sure you checkout out the svn PaperBase directory
% Open terminal
% cd <your svn path>/Marc/Longterm
% ln -s ../../PaperBase/paperbase.bib .
% ln -s ../../PaperBase/bst .
\bibliographystyle{bst/apsrev_modified_doi}
\bibliography{paperbase}
\end{document}

%% file: includes/Header.tex
\usepackage[pdftex,
        colorlinks=true,
        urlcolor=darkblue,               % \href{...}{...}
        anchorcolor=darkblue,
        filecolor=green,                 % \href*{...}
        linkcolor=darkblue,              % \ref{...} and \pageref{...}
        menucolor=darkblue,
        citecolor=darkblue,
        pagebackref,
        backref={\pdfbackref},
        pdfpagemode={UseOutlines},
        bookmarks=true,
        bookmarksopen=true,
        pdftitle={\pdftitle},
        pdfauthor={\authorname}, 
        pdfsubject={\pdfsubject},
        pdfkeywords={\pdfkeywords},
        ]{hyperref}

\usepackage{graphicx}
\DeclareGraphicsExtensions{.pdf,.png,.jpg}
\usepackage{thumbpdf}
\usepackage{color}
\input{includes/colour}

\usepackage[paper=a4paper,total={170mm,241mm},top=20mm,left=20mm,columnsep=8mm]{geometry}

\usepackage{siunitx}
\usepackage{nicefrac}

\usepackage{amsmath}
\usepackage{wasysym}
\usepackage{booktabs}

\usepackage{lineno}

\usepackage{pdfsync}

\usepackage{doi}

%% file: includes/colour.tex
\definecolor{darkgreen}{rgb}{0,0.5,0}
\definecolor{darkblue}{rgb}{0,0,0.5}
\definecolor{brown}{rgb}{0.98,0.92,0.73}
\definecolor{red}{rgb}{1,0,0}
\definecolor{yellow}{rgb}{1,1,0}
\definecolor{blue}{rgb}{0,0,1}
\definecolor{green}{rgb}{0,1,0}
\definecolor{purple}{rgb}{1,0,1}
\definecolor{gray}{rgb}{0.8,0.8,0.8}
\definecolor{black}{rgb}{0,0,0}
\definecolor{white}{rgb}{1,1,1}
\definecolor{gold}{rgb}{1.,0.84,0.}

%% file: includes/Macro.tex
\usepackage{xspace} %The \xspace command adds space at the end of a macro designed for use in text, unless the macro is followed by certain punctuation characters.
\def\um{$\mu$m\xspace}
\def\cm2{cm$^2$\xspace}

\usepackage{amsmath}

%% file: sections/Introduction.tex
\section{Introduction}
\label{sec:introduction}

When we observe motile bacteria in an optical microscope, we generally
have the problem that the bacteria are in focus for a limited period
of time. Microtechnology allows us to fabricate microfluidic channels
with a channel height of a few \si{\um}, so that the bacteria can be
forced to remain in focus (within the depth of field). The bacteria might however
still move out of the field of view. Magneto-tactic bacteria
(MTB)~\cite{Klumpp2018} offer a solution, since they can be forced to
swim along a predefined pattern by an external magnetic
field~\cite{Pichel2018}.

%\subsubsection{Research question and relevance}
In this paper, we describe an experiment in which we observed
individual MTB swimming in a figure-8 pattern for a duration of
several hours. Our question was how the velocity of the bacteria
develops over time, and what happens if the MTB stop swimming. This
question has relevance for the application of MTB as carriers for
targeted drug delivery~\cite{Felfoul2016}, in which the MTB will be
travelling for thirty minutes or more towards the tumor site. The
genes for magnetosomeformation are being identified~\cite{Uebe2018}, and it is
being investigated whether they can be expressed in other types of
microorganisms~\cite{Kolinko2014}. Novel venues can be taken for cells
acting as drug delivery agents. If cells from the human microbiota can
be genetically engineered to express magnetosomes, their lifetime in
the human body is likely to be larger than that of cells not naturally
occurring in the human body. Also the type and number of tasks that
such magnetically stearable cells can perform may increase. Long
term behaviour is also of importance for the application of MTB as
transporters inside microfluidic systems themselves~\cite{Chen2014}.

%\subsection{Prior state-of-art}

As far as we know, there have been no observations of individual
swimming bacteria over a period of more than a few minutes. MTB have
been observed in chambers fabricated from cover slides glued to
microscope slides. Reufer \emph{et al.}~\cite{Reufer2014} observed
the trajectories of MSR-1 MTB swimming along the top or bottom surface for
a few seconds. Erglis~\cite{Erglis2007} used \SI{28}{\um}-thick
double-sided tape to reduce the cell height, and observed single MSR-1
for a period up to \SI{200}{s}.

There are no reports on long time observations of individual MTB in
microfabricated chips. Other motile baceria have been observed
(\emph{E-Coli} in a channel with a height of \SI{60}{\um} \cite{Ahmed2008} and
\emph{S. marcescens} in a channel with a height of \SI{10}{\um} \cite{Binz2010}),
but observation times have been below one minute.  M\"annik \emph{et
  al.} \cite{Maennik2009} observed the growth of a culture of
\emph{E. Coli} up to two days in microfluidic chips with channel
heights of \num{5} to \SI{7}{\um}, but did not track single bacteria
for longer than a few seconds. Rather than using restriction and
magnetic bacteria, one can also move the observation cell mechanically
in the tracking microscope. To keep the bacteria in focus, the tracking
has to be done in three dimensions~\cite{Taute2015}. The longest trajectories observed were \SI{100}{s}.

We observed MSR-1 inside a glass microfabricated microfluidic chip
with a channel height of only \SI{5}{\um}, to ensure they stay in the
depth of field during the entire experiment. In combination with
magnetic control, we could observe individual MTB swimming in a
figure-8 pattern. The bacteria were observed one by one, for periods
up to \SI{70}{min}, for a total duration of the experiment of \SI{260}{min}.

%%% Local Variables:
%%% mode: latex
%%% TeX-master: "../MTBLongterm"
%%% End:

%% file: sections/Experimental.tex
\section{Experimental}
\label{sec:experimental}

\subsection{Cultivation of the magnetotactic bacteria}
\emph{Magnetospirillum gryphiswaldense} strain MSR-1 (DSM 6361) cells
were grown heterotrophically in a liquid medium (pH \num{7.0}) containing
succinate as the energy source, as previously described by Lefevre \emph{et
  al.}~\cite{Lefevre2014}. FeSO$_4$ was used as the iron source. The MSR-1 cells were
cultivated at \SI{26}{\celsius} for \num{3} to \SI{5}{days} in sterile microfuge tubes prior to microscopic analyses.

\begin{figure}
  \begin{center}
    \includegraphics[width=\linewidth]{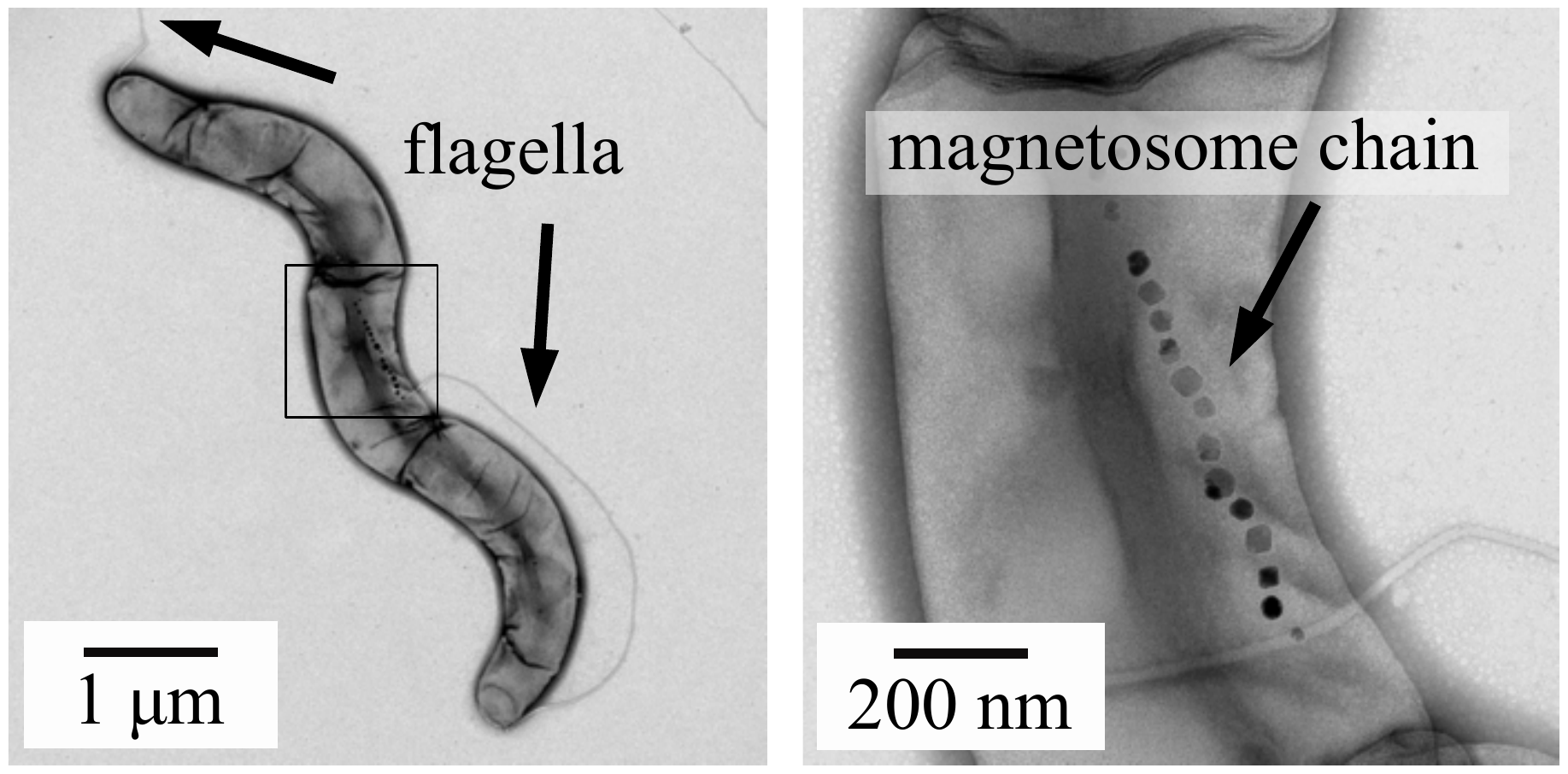}
  \end{center}
  \caption{Bright field TEM images of the
    MRS-1 magneto-tactic bacteria. In this negatively stained
    image, the flagella can be clearly observed (left) as well as the
    magnetosome chain (right).}
  \label{fig:SEM}
\end{figure}

The sampling was done using a magnetic ``racetrack''
separation method as described in~\cite{Wolfe1987}. Figure
\ref{fig:SEM} shows a transmission electron microscope (TEM) image of an
MSR-1, in which the magnetosome chain can be clearly identified.

\subsection{Transmission electron microscopy imaging}
For the TEM analysis, the medium with
MSR-1 MTB was applied on carbon coated copper grids (CF200-Cu)
and allowed to absorb for \SI{30}{sec}. Excess sample was blotted off by
touching the edge of the grid with a clean piece of filter paper and
stained with \SI{2}{\percent} uranyl acetate solution for
\SI{30}{sec}. The morphology of the MTB was examined by a JEM-2100F TEM (JEOL, Japan) with bright field image at an accelerating voltage of \SI{200}{kV}.

\subsection{Microscope}
\begin{figure}
  \begin{center}
    \includegraphics[width=.6\linewidth]{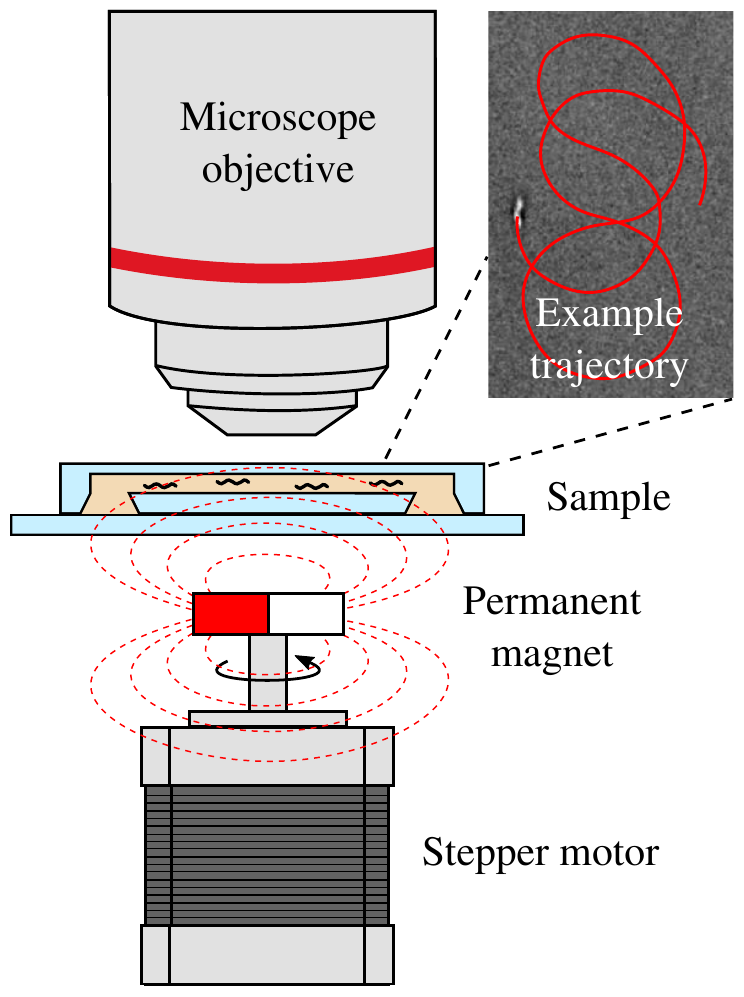}
  \end{center}
  \caption{A sample of MTB is inserted in a sealed microfluidic chip
    and observed with a reflected light microscope. A motorized
    magnet located under the sample generates in-plane magnetic
    fields, used to keep the bacteria in the field of view.}
  \label{fig:setup}
\end{figure}

Figure \ref{fig:setup} shows the experimental setup. We used an
upright reflected light microscope (Zeiss Axiotron II) with a
20$\times$ lens with a numerical aperture of \num{0.5} optimized for
reflected light, a working distance of \SI{2.1}{mm}, and field of view
of \SI{25}{mm} (Zeiss Epiplan HD DIC). Images were taken by
a CCD camera (Point Grey FL3-U3-13S2M-CS) at \SI{10}{fps} with a
resolution of 1328$\times$1024.

As light source, a collimated blue LED with an average wavelength of
\SI{470}{nm} was used (Thorlabs M470L2-C4). The manufacturer specifies
an approximate beam power of \SI{210}{mW} in a beam diameter of
\SI{37}{mm}. Upon 20$\times$ magnification, this leads to a theoretical
power density of \SI{78}{kW/m^2}. Since for imaging the aperture is
nearly closed, a large fraction of the power is blocked. By measuring
the intensity difference between a fully open and nearly closed
aperture by means of the average intensity on the camera, we estimate a
reduction by a factor of \num{30}, leading to an estimated power
density of approximately \SI{3}{kW/m^2}.

\subsection{Microfluidic Chip}
\begin{figure}
  \begin{center}
  \includegraphics[width=\columnwidth]{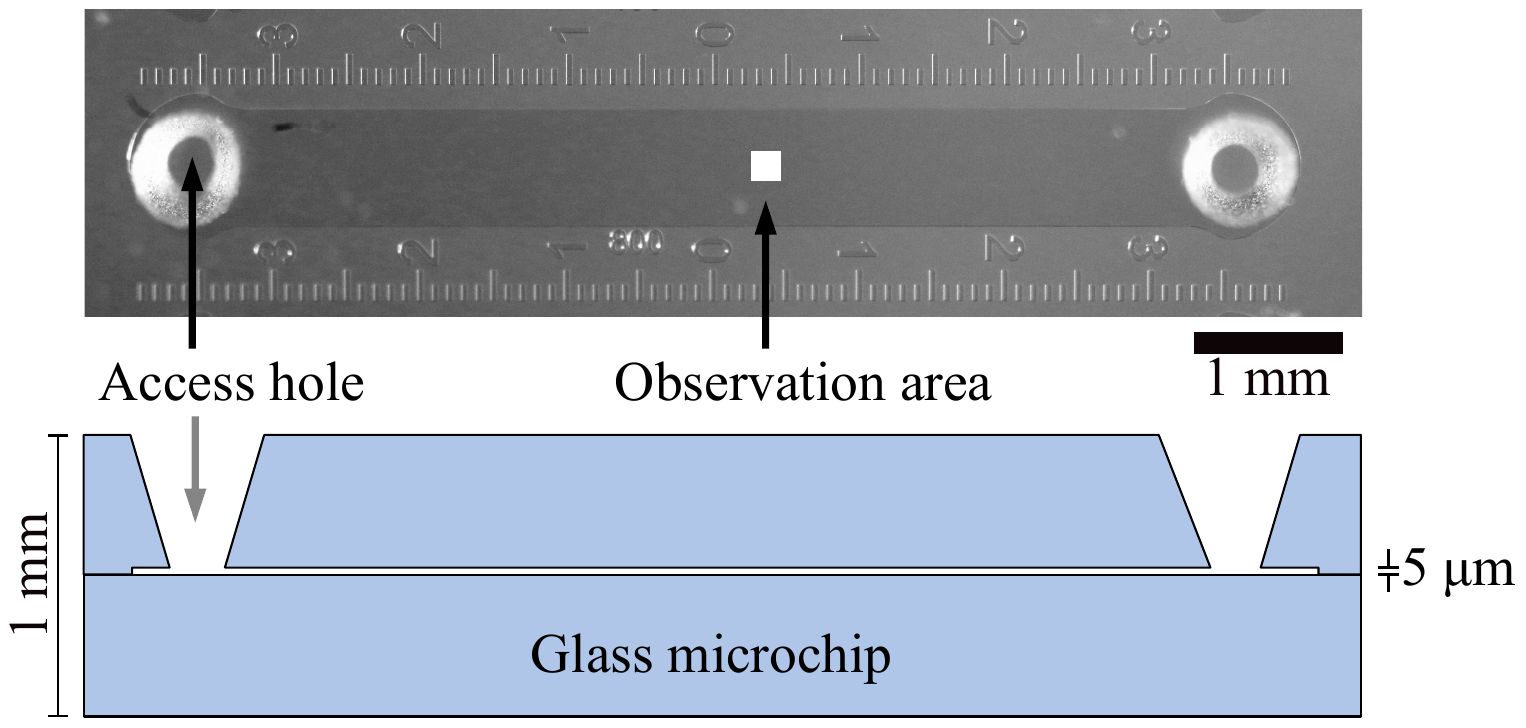}
\end{center}
\caption{A microfluidic channel with a height of only \SI{5}{\um} was
  used for the observation. MTB can be loaded through the powder
  blasted access holes. The field of view was \SI{200}{\um} (white
  box). MTB were redirected before drifting out of the centre of the
  field of view.}
\label{fig:microfluidicchip}
\end{figure}

The MTB were observed inside a microfluidic chip with a channel height
of \SI{5}{\um}, identical to our experiments in~\cite{Pichel2018}. An
overview of the microfluidic chip and region of interest for
observation can be seen in Figure~\ref{fig:microfluidicchip}.

\subsection{Magnetic field}
\begin{figure}
  \begin{center}
    \includegraphics[width=.8\linewidth]{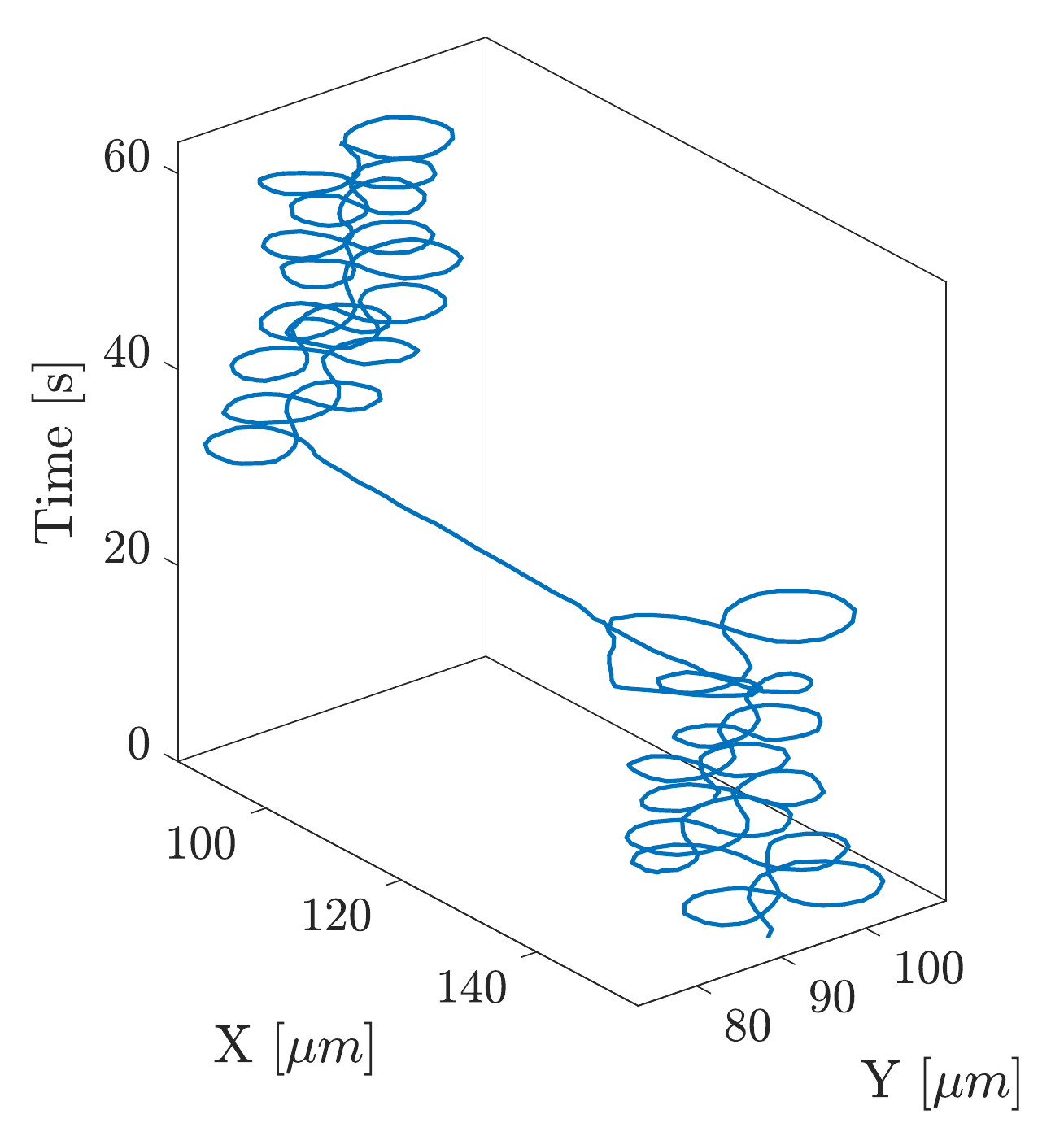}
  \end{center}
  \caption{Re-centring maneouver to keep a given MTB in the field of view. The long straight path indicates a manual correction between sequences of figure-8 loops.}
  \label{fig:recontrol}
\end{figure}

The magnetic field is generated by a motorised permanent magnet placed
underneath the sample~\cite{Pichel2018}. This magnet has its
magnetisation orthogonal to the axis of rotation, so that it creates
an in-plane magnetic field at the location of the sample. The in-plane
angle of the field can be controlled by the rotation of the motor
axis. The motor was programmed to loop in a figure-8 trajectory, so
that the bacteria, on average, will not change their position. The
programmed trajectory can be manually overridden, to steer the
bacteria back to the centre of view to correct for drift.  The
tracking results of a typical re-centring process can be seen in
Figure~\ref{fig:recontrol}.

\subsection{Image Processing}
The image sequence was processed offline to extract the coordinates of
the bacteria of interest. The procedure is identical to the method
published in~\cite{Pichel2018}. The low-contrast nature of the image
required pre-processing steps. Subsequently, we performed background
subtraction, lowpass filtering, thresholding, and finally selecting the
resulting blobs based on size. The centre of gravity was registered as
the position of the bacteria. A nearest-neighbour algorithm with maximum
search radius was used to build the trajectories from the detected
bacteria. The resulting trajectories were manually cleaned for
drop-outs. The velocity was calculated from the trajectories. Due to
noise, the centre of gravity of the blobs jitters by one or two pixels
(\SI{180}{nm} per pixel). As the velocity is calculated from the
frame-by-frame displacement at \SI{10}{fps}, this will result in a
small residual velocity of approximately~\SI{3}{\um/s}.

% \cmtr{Matlab code, Tijmen.}

%%% Local Variables:
%%% mode: latex
%%% TeX-master: "../MTBLongterm"
%%% End:

%% file: sections/Results.tex
\section{Results}
\label{sec:results}

\subsection{Long term tracking}

Data was recorded for a period of \SI{5}{\hour}. During this period, a
single MTB was tracked at a time. An example of a recorded video is
available as additional material (figure8.mov). When the selected MTB
stopped moving, the magnetic field was directed to be parallel to the
microfluidic channel in order to find and trap a new MTB.

\begin{figure}
  \begin{center}
    \includegraphics[width=.9\linewidth]{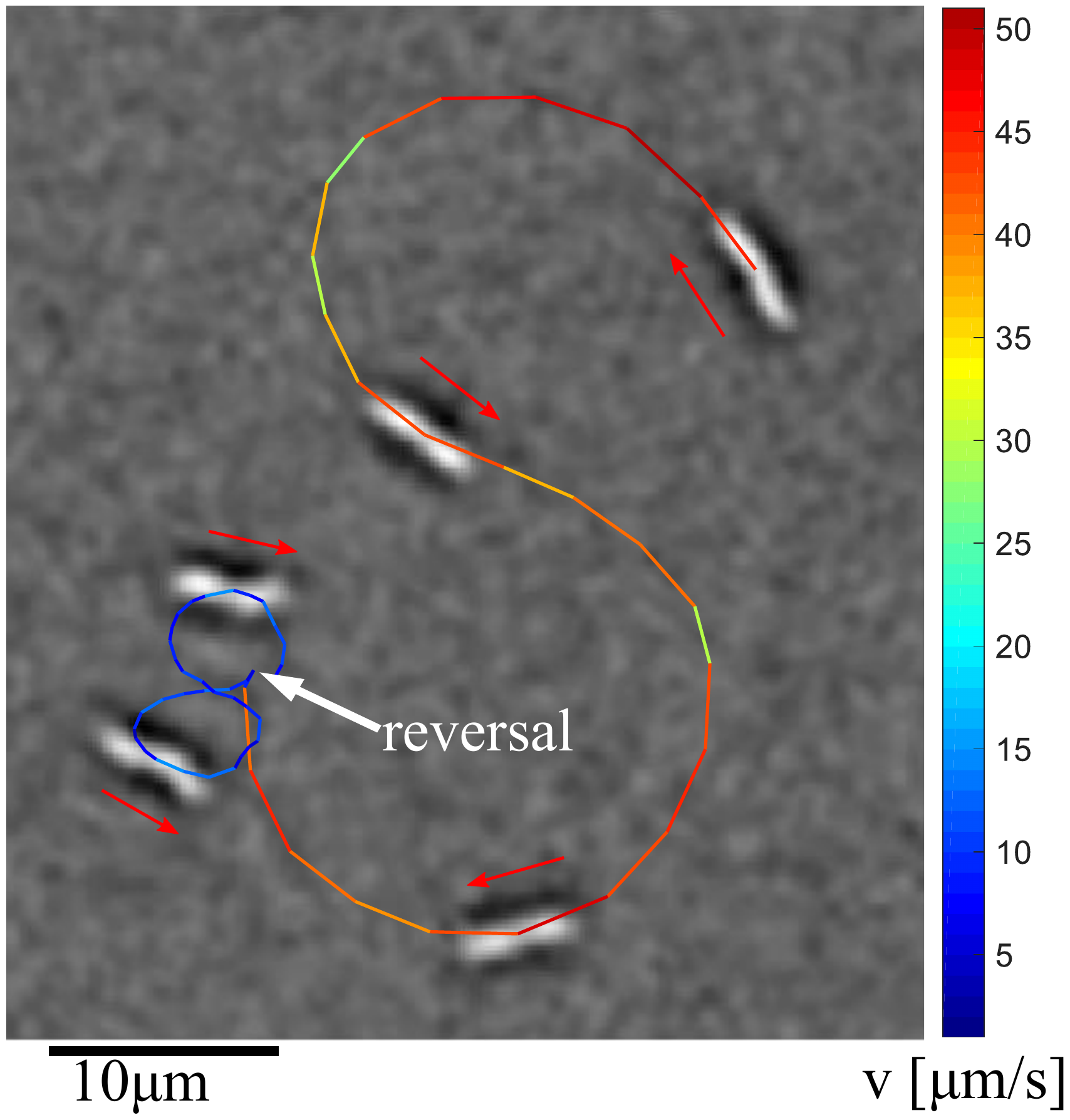}
  \end{center}
  \caption{Two figure-8 trajectories of a single MTB. At the start of
    the observation, the MTB travels at a velocity of \num{30} to
    \SI{50}{\um/s} resulting in a big trajectory. After
    \SI{2.4}{s} the MTB reverses direction and its velocity
    drops to \SI{5}{\um/s}, resulting in a much smaller
    trajectory.}
  \label{fig:reversal}
\end{figure}

Figure~\ref{fig:reversal} shows a composite image of one single
MTB. The total trajectory with a length of \SI{6}{s} is shown. On top
of the trajectory five frames of the recorded movie are
superimposed. At the start of the observation, the MTB follows big
figure-8 shaped trajectories at a velocity of up to \SI{50}{\um/s}. After
\SI{2.4}{s}, the MTB reverses direction and continues to swim at a
very low speed of \SI{5}{\um/s}, resulting in a small figure-8 shaped
trajectory. This behaviour is in agreement with the bimodal velocity
distribution previously observed by Reufer \emph{et al.}~\cite{Reufer2014}.

\begin{figure}
  \centering
  \includegraphics[width=\columnwidth]{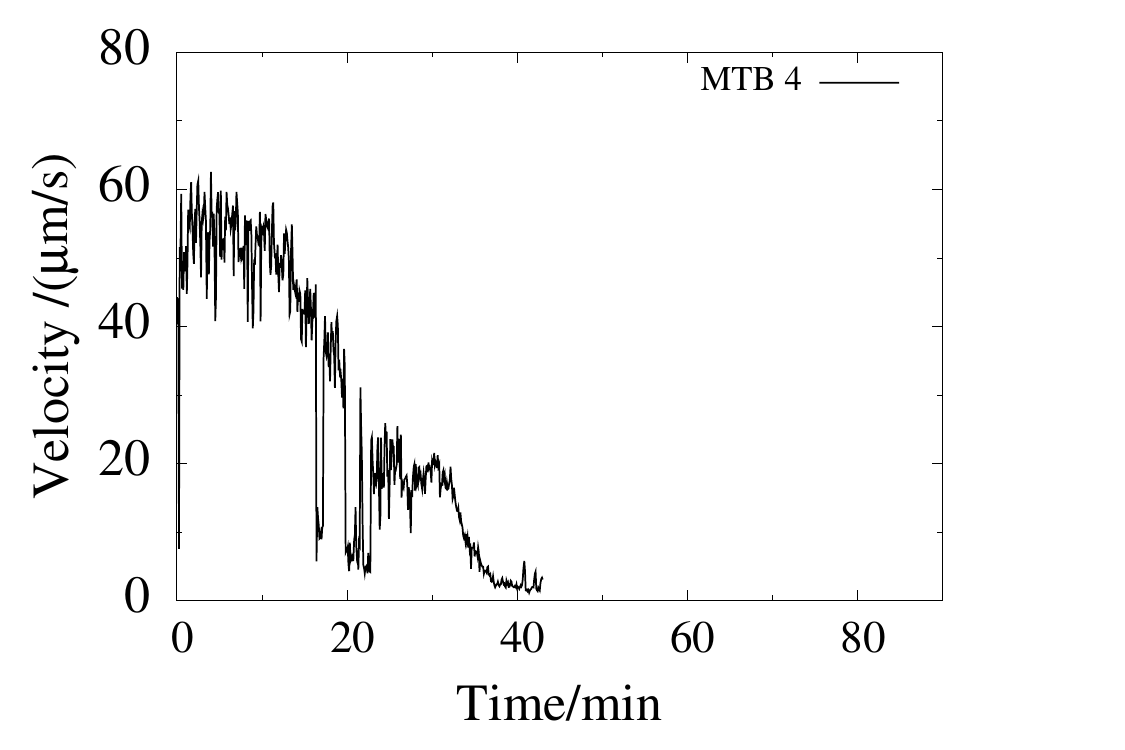}
  \includegraphics[width=\columnwidth]{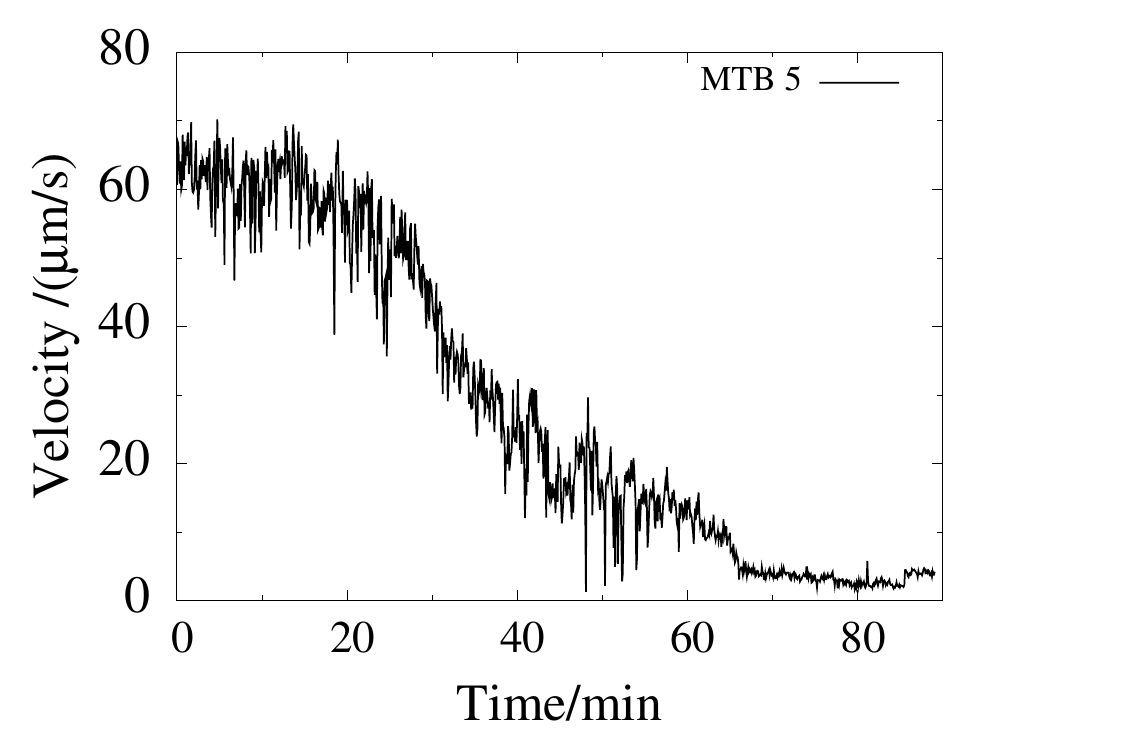}
  \caption{Velocity of two individual MTB versus time. The initial
    velocity is \num{50} to \SI{60}{\um/s}. The velocity decreases
    with time over a period of about half an hour, until the MTB stop
    moving.}
   \label{fig:data}
\end{figure}

Figure~\ref{fig:data} shows the velocity of two MTB as functions of
time.  The MTB initially show a constant velocity, after which the
velocity gradually decreases with time. The initial velocity and the
duration of the period of constant velocity vary. For MTB4, which is
the same MTB as in Figure~\ref{fig:reversal}, several
reversals can be observed between \num{18} and \SI{22}{min}.

\begin{figure}
  \centering
  \includegraphics[width=\columnwidth]{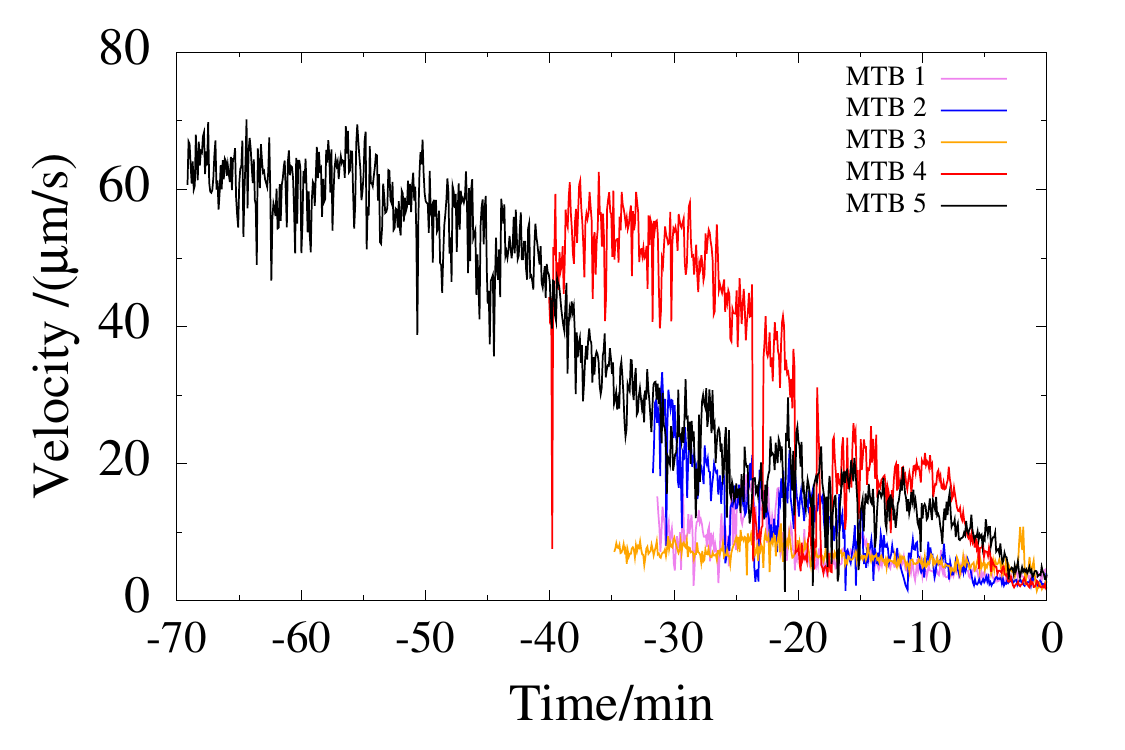}
  \caption{Velocity of observed MTB as a function of time. The
    time at which the bacteria stops moving is taken as
    reference ($t$=0). This way of displaying clearly suggests that
    the decrease in speed shows a similar behaviour between the
    MTB.}
   \label{fig:combined}
\end{figure}

There appears to be more or less a similar rate, about \SI{1.5}{\um/s}
per minute (\SI{25}{nm/s^2}), in the decrease of velocity. Therefore we
plotted the velocity of all observed MTB as a function of time, taking
the point at which the bacteria stops moving forward as a reference,
see Figure~\ref{fig:combined}. The figure suggests that we captured
the full behaviour of MTB4 and MTB5, but observed the other MTB at the
end of the decay process. This is most likely because we became more
skilled in capturing new MTB over the course of the \SI{5}{h}
experiment.

\subsection{Modes of motile behaviour}
\begin{figure}
  \includegraphics[width=.3\linewidth]{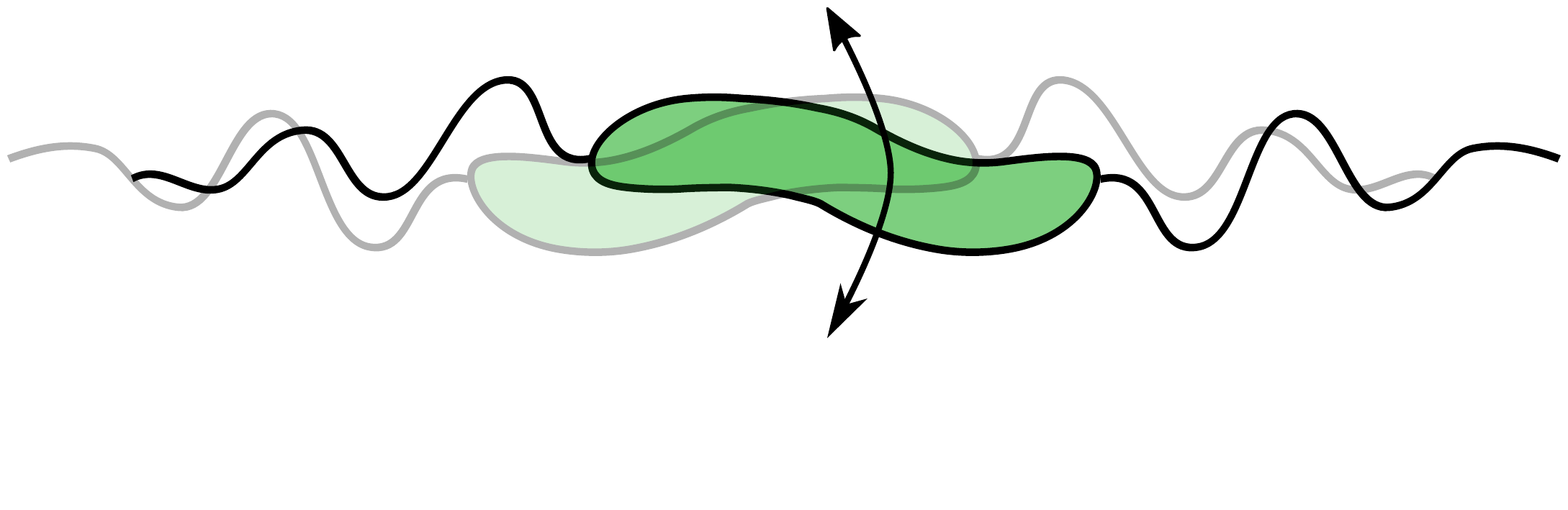}
  \includegraphics[width=.65
  \linewidth]{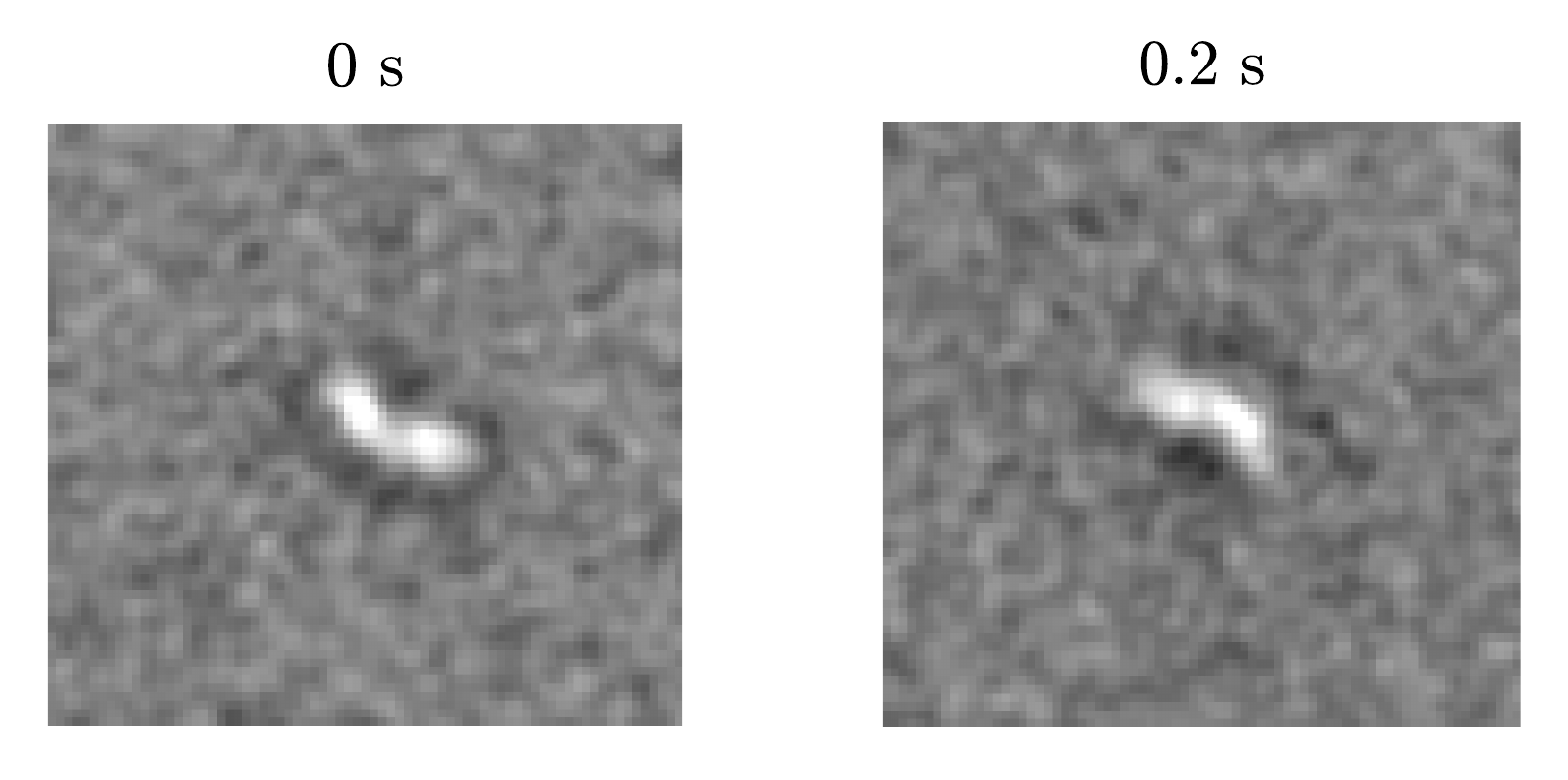}
  \caption{An image sequence of an MTB rotating around the long axis.}
  \label{fig:rot_minor}
\end{figure}

When the MTB slow down, we can observe a rotation around the long
axis of the MTB (Figure~\ref{fig:rot_minor}), which is in agreement
with propulsion by a rotating flagellum~\cite{Purcell1977}. 

\begin{figure}
  \includegraphics[width=.3\linewidth]{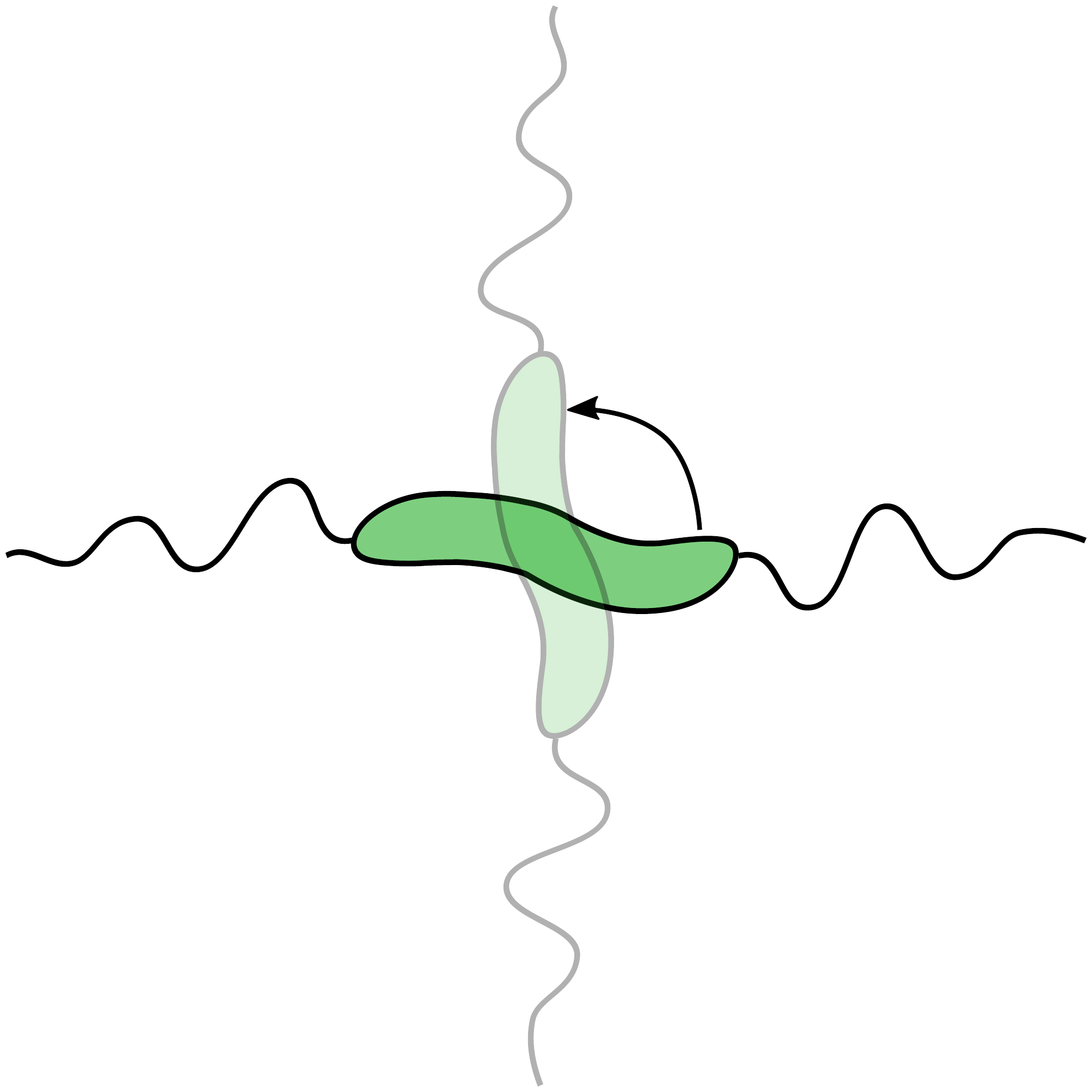}
  \includegraphics[width=.65\linewidth]{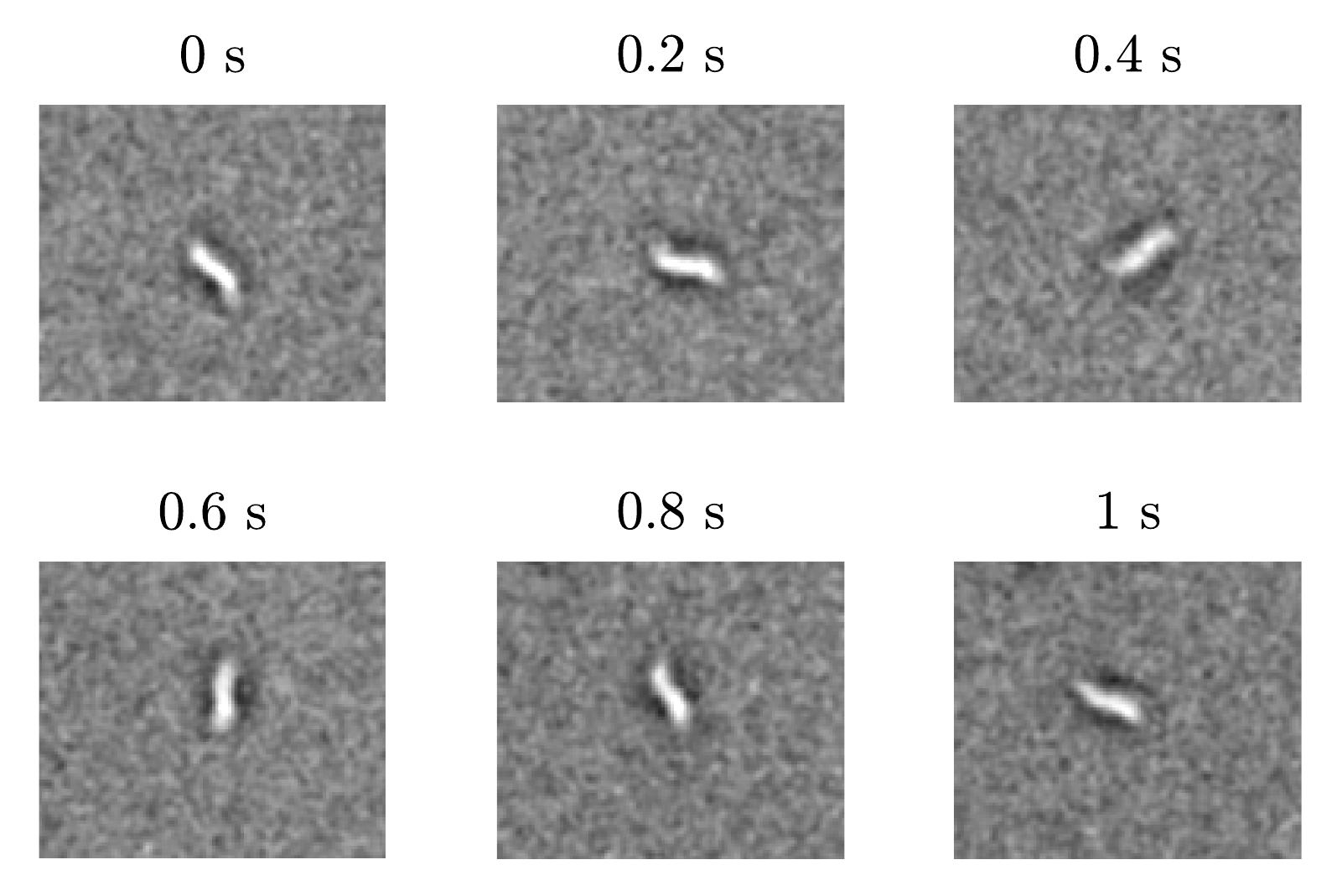}
  \caption{Even if an MTB no longer moves forward, it still is
    rotating due
    to the torque generated by the rotating external field. As a
    result, even non-motile MTB rotate around an axis perpendicular to
    their body.}
  \label{fig:rot_major}
\end{figure}

\begin{figure}
  \includegraphics[width=.3\columnwidth]{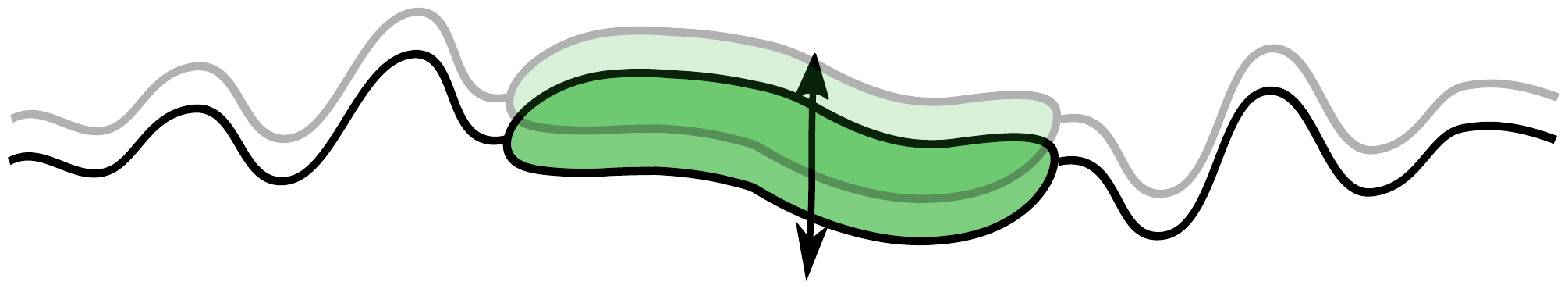}
  \includegraphics[width=.65\columnwidth]{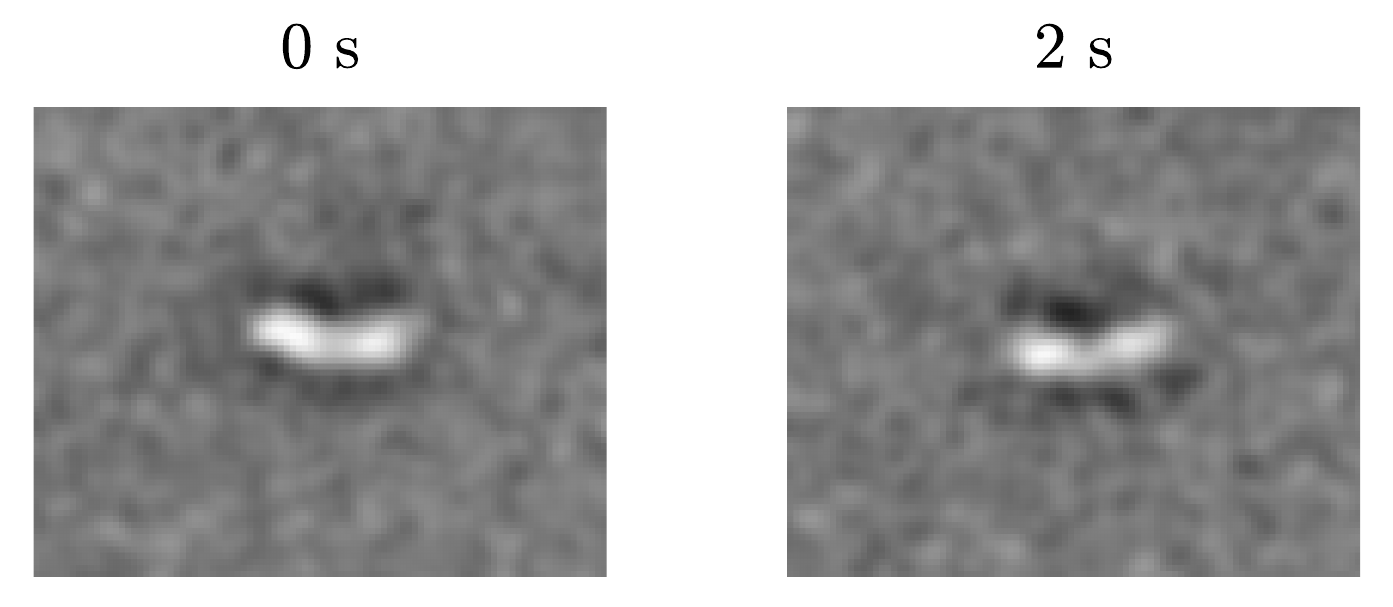}
  \caption{If an MTB does not rotate in the field at all but shows
    Brownian motion, it can be assumed that it does not have a
    magnetosome.}
  \label{fig:stop}
\end{figure}

When the MTB stop swimming, we can still observe movement. There is a
difference between MTB that have a magnetosome, and those that do
not. The external rotating field will always exert a torque on an MTB
with a magnetosome, even if it is dead. We observe these MTB rotating
around an axis that appears to be very close to the centre of their
body (Figure~\ref{fig:rot_major}). We also observe MTB that are not
rotating at all (Figure~\ref{fig:stop}). These would be either MTB
without a magnesome, or MTB that are firmly stuck to 
the channel wall. Since we observe small random motion, in agreement
with Brownian motion, we assume they are non-magnetic.

\begin{figure}
  \includegraphics[width=.3\linewidth]{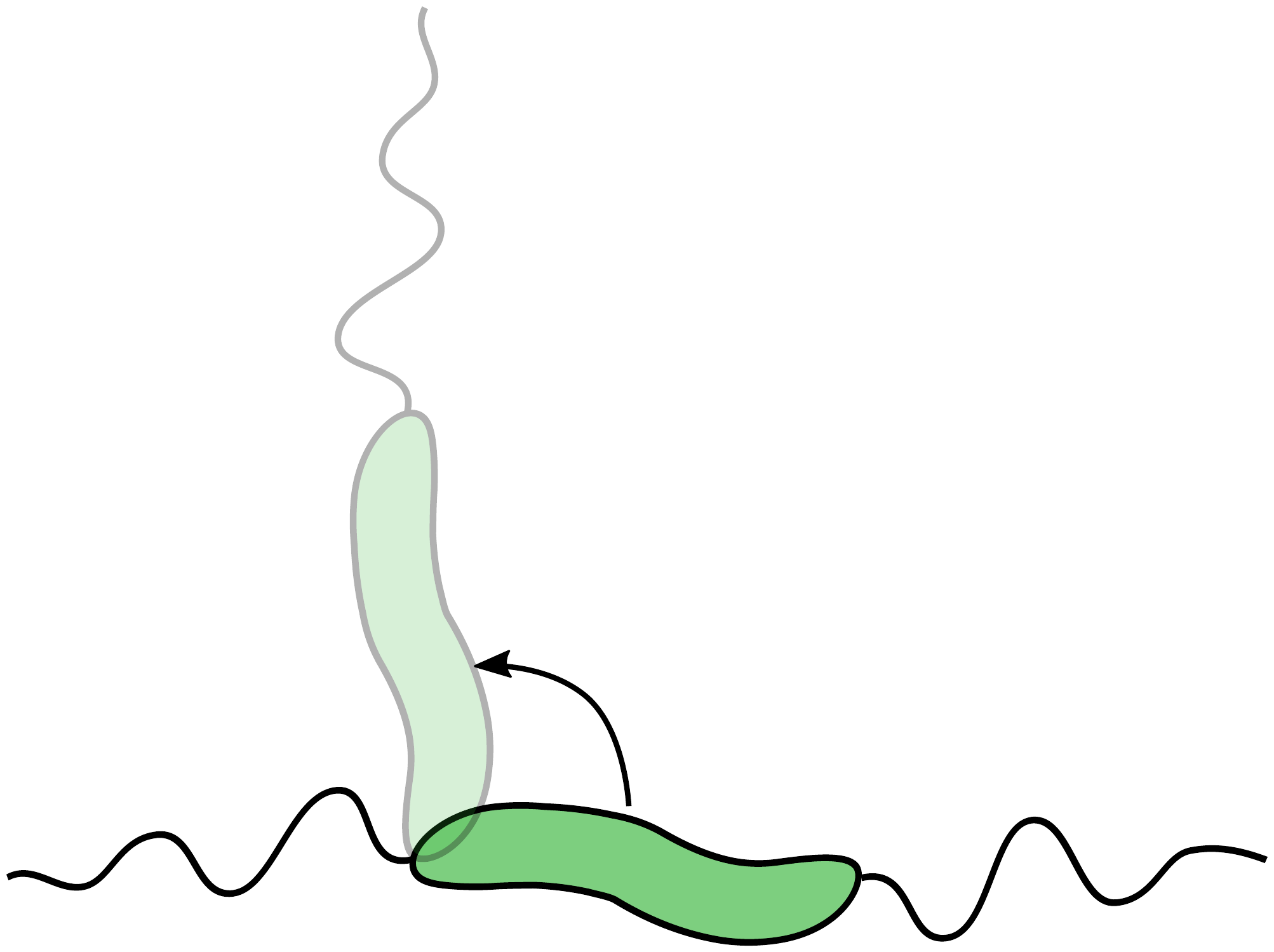}
  \includegraphics[width=.65\linewidth]{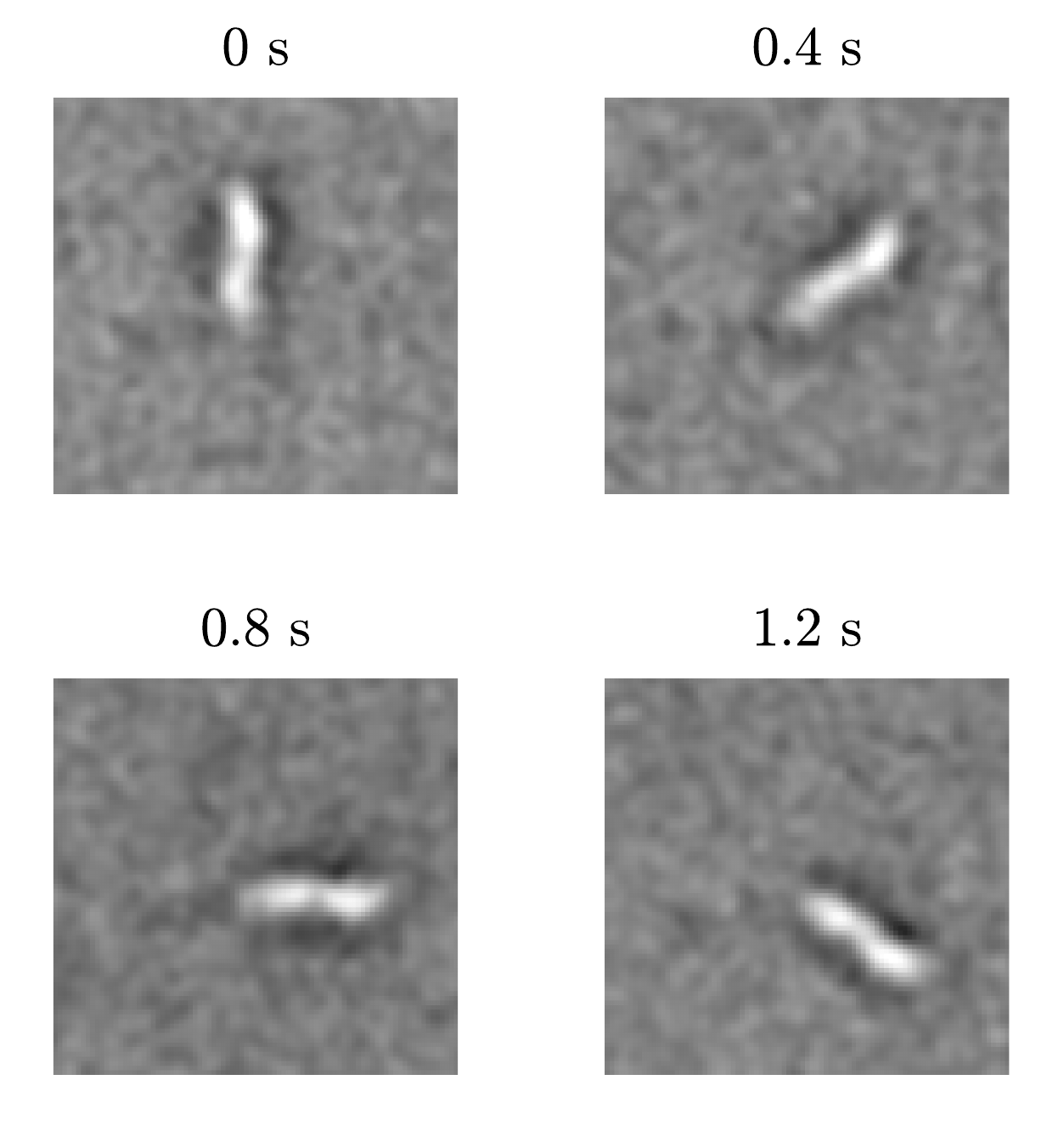}
  \caption{Magnetic and non-magnetic MTB sometimes rotate around one
    end.  These are probably stuck to the channel wall with a flagellum.} \label{fig:rot_point}
\end{figure}

There are other MTB that appear to be stuck to the surface of the
channel. They rotate around a point that is not in the centre of their
body (Figure~\ref{fig:rot_point}). MTB with and without magnetosome displayed
this behaviour. Magnetic MTB follow the rotation of the field but
non-magnetic MTB rotate randomly.

%%% Local Variables:
%%% mode: latex
%%% TeX-master: "../MTBLongterm"
%%% End:

%% file: sections/Discussion.tex
\section{Discussion}
\label{sec:discussion}

%\subsection{Decline in velocity}
In each case of a long term control sequence we observe a decline in
velocity. There a several hypotheses.

% We koken de bacterieen...
A possible explanation might be that the direct lighting from our
observation is heating the samples we are observing to a
temperature that kills the MTB. However, no noticeable increase in chip
temperature was observed. Since the thermal conductivity of
glass is high, it does not support a steep thermal gradient, so the
entire chip would be at the same temperature. We were still able to find
new motile MTB during a \SI{5}{hour} experiment. It is therefore
  unlikely that the observation area increases significantly in
  temperature.

% Ze houden er uit zichzelf mee op:
% Honger
We assume that local ion depletion is also not a problem, since in
some cases when an MTB comes to a halt, other non-magnetic ones still
cross the screen without a problem.

% Tijd voor sex
One could imagine MTB stop swimming prior to cell division. Clear
size changes however are not visible, even after observing the
bacteria for a long period after they stop. From growth curves, we can
estimate the mean time between cell division to be on the order of
\num{4}-\SI{8}{hours}~\cite{Sun2008}. The chance that all observed MTB
stopped swimming because there were close to cell division is
negligble (less than \SI{0.1}{\percent}).

% Moe
One might suggest the reduction in velocity is simply due to fatigue
of some sort. Perhaps the flagellar motor has intervals of activity
over such a long period of time.  The efficiency of the flagellar
motors is reported to be near
\SI{100}{\percent}~\cite{Kinosita2000}. Moreover, we have never observed
an MTB starting to move again after coming to a halt.

% Te veel licht
The most likely explanation is an overdose of light. In the
microscope, we focus a very bright LED light souce of \SI{450}{nm}
wavelength on the field of view. This high intensity light source
might damage the bacteria under observation. It is known that
magneto-tactic bacteria respond to light. For MC-1 bacteria,
illumination seems to have the same effect as an increase in oxygen
concentration~\cite{Frankel1997}. Non-magnetic AMB-1 bacteria have
been observed to migrate towards the light~\cite{Li2017}. There are
no published reports on photo-toxicity of magneto-tactic bacteria, but
we could assume a similar sensitivity to light as other types of
bacteria. Light illumination at \SI{415}{nm} at a dose of
\SI{750}{kJ/m^2} for instance strongly reduces the viability of
\emph{Propionibacterium acnes}
cultures~\cite{Ashkenazi2003}. Similarly, Santos \emph{et
  al.}~\cite{Santos2013} showed that a UV-light dose of
\SI{300}{kJ/m^2} at \SI{365}{nm} wavelength is sufficient to reduce
the viability of most of a set of nine different types of surface
water bacteria. Assuming an illumination power density at the sample
plane of \SI{3}{kW/m^2} (see Section~\ref{sec:experimental}), this
would be equivalent to an exposure time of less than \SI{250}{s}.

This photo-toxic explanation is also in agreement with the observation
that we can always find a new motile MTB. Only the MTB that are within
the illumnated area (about \SI{1.2}{mm^2} will suffer from the intense
light source. The MTB outside this area remain unaffected
until we use them for observation. It is therefore reasonable to
assume that indeed an overdose of light is responsible for the decrease in
MTB motility over time.

%% Local Variables:
%% mode: latex
%% TeX-master: "MTBLongterm"
%% End:

%% file: sections/Conclusion.tex
\section{Conclusion}
\label{sec:conclusion}

% Onderzoeksvraag
%What can we learn from observing for an extended amount of time?

We observed magneto-tactic bacteria of type \mbox{MSR-1} inside a
microfluidic chip for a total of \SI{260}{\minute}. During this time,
individual bacteria were magnetically steered in figure-8 patterns for
a duration of from twenty to fifty minutes.

The MTB occasionally reverse direction, which is accompanied with a
sudden drop in velocity. All observed bacteria showed a gradual
decrease in velocity until they came to a full stop. The time until
the start of the decrease varied, with a maximum of \SI{30}{\min}. The
decay rate however was relatively constant, at about
\SI{25}{nm/s^2}.

When the MTB slow down but are still swimming, we can observe rotation
around their long axis. After coming to a halt, we observed three
different behaviors. 1) Many MTB still rotate in the field around an
axis perpendicular to their long axis and close to their centre of
mass. 2) Some MTB appear to be stuck to the channel wall with a
flagellum and rotate around one end. They either rotate synchronously with
the rotation of the field or randomly. The latter group must be
non-magnetic. 3) Finally there are MTB that do not move at all, either
because they are firmly stuck or non-magnetic.

As far as we know, this experiment is the first observation of individual motile
bacteria for an extended period of time. The experiment was enabled by
the availability of glass microfabricated chips with low channel
height, so that the bacteria stay in focus. We learned from the
experiment that one should consider the influence of the microscope
light and the presence of the channel walls.

%%% Local Variables:
%%% mode: latex
%%% TeX-master: "..\MTBLongterm"
%%% End: